\documentclass[journal,10pt]{IEEEtran}

\usepackage{cite}
\usepackage{graphicx}
\usepackage{amsmath}
\usepackage{array}
\usepackage{mdwmath}
\usepackage{amssymb}
\usepackage{mdwtab}
\usepackage{stfloats}
\usepackage[tight,footnotesize]{subfigure}
\usepackage{amsmath,amsthm}
\usepackage{threeparttable}
\usepackage{color}
\usepackage{url}
\usepackage{algpseudocode}
\usepackage{algorithm}
\usepackage{multicol}
\usepackage{amssymb}
\usepackage{flushend}

\algrenewcommand{\algorithmicrequire}{\textbf{Input:}}
\algrenewcommand{\algorithmicensure}{\textbf{Output:}}

\ifCLASSINFOpdf
\else
\fi

\begin{document}

\title{Quadrature Spatial Scattering Modulation for mmWave Transmission}
\author{ Xusheng Zhu, Wen Chen, \IEEEmembership{Senior Member, IEEE}, Zhendong Li, \\  Qingqing Wu, \IEEEmembership{Senior Member, IEEE}, and Jun Li, \IEEEmembership{Senior Member, IEEE}
\thanks{This work is supported by National key project 2020YFB1807700, NSFC 62071296, Shanghai 22JC1404000, 20JC1416502, and PKX2021-D02. (\emph{Corresponding author: Wen Chen}.)}
\thanks{Xusheng Zhu, Wen Chen, Zhendong Li, and Qingqing Wu are with the Department of Electronic Engineering, Shanghai Jiao Tong University, Shanghai 200240, China (e-mail: xushengzhu@sjtu.edu.cn; wenchen@sjtu.edu.cn; lizhendong@sjtu.edu.cn; qingqingwu@sjtu.edu.cn).}
\thanks{Jun Li is with the School of Electronic and Optical Engineering,
Nanjing University of Science Technology, Nanjing 210094, China (e-mail:
jun.li@njust.edu.cn).}
}

\maketitle

\begin{abstract}
In this letter, we investigate a novel  quadrature spatial scattering modulation (QSSM) transmission technique based on millimeter wave (mmWave) systems, in which the transmitter generates two orthogonal beams targeting candidate scatterers in the channel to carry the real and imaginary parts of the conventional signal, respectively.
Meanwhile, the maximum likelihood (ML) detector is adopted at the receiver to recover the received beams and signals.
Based on ML detector, we derive the closed-form average bit error probability (ABEP) expression of the QSSM scheme.
Furthermore, we evaluate asymptotic ABEP expression of the proposed scheme.
Monte Carlo simulations verify the exactness and tightness of the derivation results.
It is shown that the ABEP performance of QSSM is better than that of traditional spatial scattering modulation.
\end{abstract}
\begin{IEEEkeywords}
Quadrature spatial scattering modulation (QSSM), millimeter wave (mmWave), maximum likelihood (ML), average bit error probability (ABEP).
\end{IEEEkeywords}

\section{Introduction}

Spatial modulation (SM), which activates only one transmit antenna for each transmission, has been considered a potential multi-antenna solution for wireless communication systems \cite{mesl2008spatial,wen2019a}. To clarify the transmission characteristics of SM in the real-world environment, the authors of \cite{zhu2022on} studied the average bit error probability (ABEP) performance of three-dimensional SM over the measured channels. To further improve the spectral efficiency of SM, quadrature SM (QSM) was proposed, in which the spatial symbol is expanded into two orthogonal in-phase, and quadrature parts \cite{meslen2015quadr}.
Since the two transmitted data are orthogonal and modulated by the orthogonal carriers, inter-channel interference (ICI) can be eliminated \cite{li2021single,li2017gen,li2021gen}.

Due to the limited spectrum resources, communications in the millimeter wave (mmWave) band are attracting widespread attention.
To combat the severe path loss, it is necessary to utilize multi-antenna techniques
\cite{sun2014mimo}.
In this manner, the gain provided by beamforming can compensate for the attenuation of mmWave signals, ensuring the stability and reliability of the data transmission.
Motivated by this, \cite{ding2017ssm} introduced spatial scattering modulation (SSM) scheme for mmWave communication, where transmitter sends a narrow beam aimed at the target scatterer, instead of activating a transmit antenna in the SM scheme.
In this context,
adaptive SSM (ASSM), generalized SSM (GSSM), and reconfigurable intelligent surface-assisted SSM (RIS-SSM) schemes were studied in \cite{zhang2021adap,zhang2022gen,zhu2022recon}.
It is worth mentioning that the SSM technique can be perfectly applied to secure communications, since the passive scatterers themselves do not carry information.

\begin{figure*}[t]
  \centering
  \includegraphics[width=16cm]{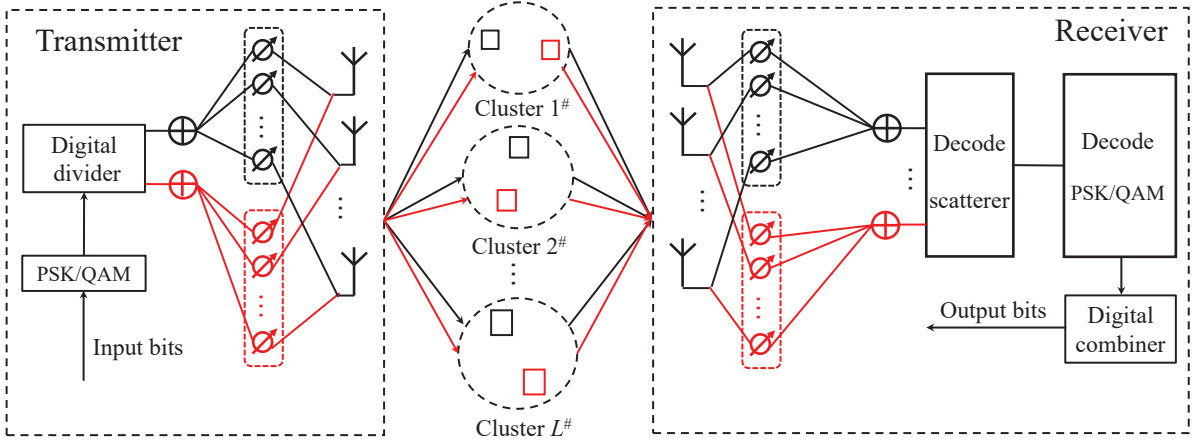}\\
  \caption{System model of the proposed QSSM scheme.}\label{systemmod}
\end{figure*}

In the SSM scheme, the real and imaginary parts of the phase shift keying/quadrature amplitude modulation (PSK/QAM) signal are emitted by a single beam. To enhance spectral efficiency, we propose a novel quadrature SSM (QSSM) scheme that preserves nearly all the characteristics of SSM.
In QSSM, the signal of the symbol domain is extended into real and imaginary parts carried by quadrature and in-phase beams, respectively.
To the best of our knowledge, research in this area is still missing in open work.
For clarity, the main contributions of this letter are summarized as follows:
\begin{itemize}
\item A novel QSSM scheme is proposed to improve the spectral efficiency for mmWave transmission.
\item Based on the maximum likelihood (ML) detector, we derive the closed-form analytical ABEP and asymptotic ABEP expressions for the QSSM scheme, respectively.
\item
The simulation results validate the analysis and demonstrate that QSSM transmission is superior to conventional SSM in terms of reliability.
\end{itemize}

\section{System Model}
In this section, we introduce the $N_r \times N_t$ multiple-input multiple-output (MIMO) model in Fig. \ref{systemmod}.
It is assumed that both transmitter and receiver can obtain perfect channel state information (CSI) \cite{alk2014channel}.
Moreover, the narrowband block-fading propagation channel is considered in this letter \cite{aya2014spati,ding2017ssm}.

\begin{table}[t]
\renewcommand\arraystretch{1.5}
\centering
\caption{\small{Bit mapping of QSSM for $M = 4$, $L = 4$.}}
\begin{tabular}{c|c|c|c}
\hline \hline$\left[b_{1} b_{2} b_{3} b_{4} \right]$ & {{[${l_{\Re}}$, ${l_{\Im}}$]}} & {$\left[b_{1} b_{2} b_{3} b_{4}\right]$} & {[${l_{\Re}}$, ${l_{\Im}}$]} \\
\hline
0 \ 0 \ 0 \ 0 & $1^\#$, $1^\#$ & 1 \ 0 \ 0 \ 0 & $3^\#$, $1^\#$ \\
0 \ 0 \ 0 \ 1 & $1^\#$, $2^\#$ & 1 \ 0 \ 0 \ 1 & $3^\#$, $2^\#$ \\
0 \ 0 \ 1 \ 0 & $1^\#$, $3^\#$ & 1 \ 0 \ 1 \ 0 & $3^\#$, $3^\#$ \\
0 \ 0 \ 1 \ 1 & $1^\#$, $4^\#$ & 1 \ 0 \ 1 \ 1 & $3^\#$, $4^\#$ \\
0 \ 1 \ 0 \ 0 & $2^\#$, $1^\#$ & 1 \ 1 \ 0 \ 0 & $4^\#$, $1^\#$ \\
0 \ 1 \ 0 \ 1 & $2^\#$, $2^\#$ & 1 \ 1 \ 0 \ 1 & $4^\#$, $2^\#$ \\
0 \ 1 \ 1 \ 0 & $2^\#$, $3^\#$ & 1 \ 1 \ 1 \ 0 & $4^\#$, $3^\#$ \\
0 \ 1 \ 1 \ 1 & $2^\#$, $4^\#$ & 1 \ 1 \ 1 \ 1 & $4^\#$, $4^\#$ \\
\hline \hline$\left[b_{5} b_{6}\right]$ & $[x_{\Re},x_{\Im}]$ & {$\left[b_{5} b_{6}\right]$} & ${[x_{\Re},x_{\Im}]}$ \\
\hline 0 \ 0 & ${\frac{-1}{\sqrt{2}},\frac{-1}{\sqrt{2}}}$ & 1 \ 1 & $\frac{1}{\sqrt{2}},\frac{1}{\sqrt{2}}$ \\
0 \ 1 & ${\frac{-1}{\sqrt{2}},\frac{1}{\sqrt{2}}}$ & 1 \ 0 & ${\frac{1}{\sqrt{2}},\frac{-1}{\sqrt{2}}}$ \\
\hline \hline
\end{tabular}
\vspace{-10pt}
\end{table}

\subsection{Channel Model}

In this subsection, we employ a geometric Saleh-Valenzuela channel model to charctrize the channel of the proposed QSSM systems.
As shown in Fig. 1, we assume that there are $L$ clusters between the receiver and the transmitter. Then, we select the two scatterers with the weakest correlation in each cluster according to the perfect CSI.
Therefore, the channel model of the QSSM scheme can be expressed as
\begin{equation}
\mathbf{H} = \sum\limits_{{k_1}=1}^{L} \beta_{k_1} \boldsymbol{a}_r(\theta_{k_1}^r)\boldsymbol{a}_t^H(\theta_{k_1}^t)
+\sum\limits_{{k_2}=1}^{L} \beta_{k_2} \boldsymbol{a}_r(\theta_{k_2}^r)\boldsymbol{a}_t^H(\theta_{k_2}^t),
\end{equation}
where $(\cdot)^H$ represents the conjugate transposition and $\beta_l$ is the complex gain of the $l$-th scatterer, i.e., $\beta_l \sim \mathcal{CN}(0,1), \forall l$, where $\mathcal{CN}(\cdot,\cdot)$ denotes circularly symmetric complex Gaussian distribution.
The antenna array response vectors $\boldsymbol{a}_r(\theta_{l}^r) \in \mathbb{C}^{N_r \times 1}$ and $\boldsymbol{a}_t(\theta_{l}^t) \in \mathbb{C}^{N_t \times 1}$
can be respectively expressed as
\begin{equation}
\begin{aligned}
\boldsymbol{a}_r(\theta_l^r) & =\!
\frac{1}{\sqrt{N_r}}[1,\exp(j2\pi\phi_l^r),\ldots,\exp(j2\pi\phi_l^r\!\cdot\!(N_r-1))]^T,\\
\boldsymbol{a}_t(\theta_l^t) & =\! \frac{1}{\sqrt{N_t}}[1,\exp(j2\pi\phi_l^t),\ldots,\exp(j2\pi\phi_l^t\!\cdot\!(N_t-1))]^T,
\end{aligned}
\end{equation}
where $(\cdot)^T$ is the transpose operator, $\phi_l^r={ d_r\sin(\theta_l^r)}/{\lambda}$,
and $\phi_l^t={ d_t\sin(\theta_l^t)}/{\lambda}$.
Further, $\theta_l^r \in [0, 2\pi]$ and $\theta_l^t \in [0, 2\pi]$ are the $l$th scatterer's angle-of-arrival (AoA) and angle-of-departure (AoD) of the receiver and transmitter, respectively.
Furthermore, $\lambda$ denotes the signal wavelength, $d_r$ and $d_t$ represent the distance between adjacent elements in the receive antenna array and the transmit antenna array, respectively.

Note that when $N_r$ and $N_t$ are large, the resulting beam is very narrow. At this point, we consider that there is no interference between the beams\cite{zhu2022recon,ding2017ssm,zhang2021adap,zhang2022gen}
\begin{equation}\label{zhenj}
\begin{aligned}
&\boldsymbol{a}_r^H(\theta_l^r)\boldsymbol{a}_r(\theta_{l'}^r) \approx \delta(l-l'), \ \ l, {l'} \in \{1,2,\ldots, L\}, \\
&\boldsymbol{a}_t^H(\theta_l^t)\boldsymbol{a}_t(\theta_{l'}^t) \approx \delta(l-{l'}), \ \ l, {l'} \in \{1,2,\ldots, L\},
\end{aligned}
\end{equation}
where $\delta(\cdot)$ denotes the Dirac delta function.

\begin{figure}[t]
  \centering
  \includegraphics[width=9cm]{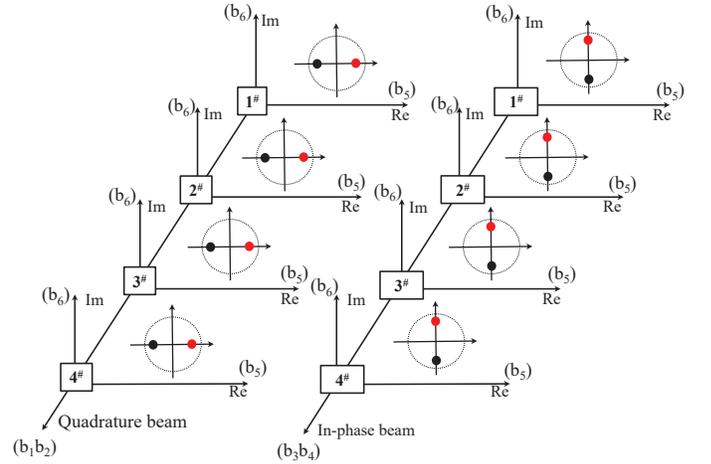}\\
  \caption{Constellation diagram with 4QAM and 4 scatterers.}\label{constel}
\end{figure}

\subsection{QSSM Transmission}

In Fig. \ref{systemmod}, the input bits are first modulated by $M$-ary PSK/QAM signal. Then, the modulated signal is decomposed into real and imaginary parts. Next, the transmitter sends quadrature and in-phase beams to carry real and imaginary parts of the signal, respectively.
Particularly, two orthogonal beams can activate two candidate scatterers in the scattering cluster. As a result, the spectral efficiency of the QSSM system can be given by
\begin{equation}
R = \log_2(M) + \log_2(L) + \log_2(L) \ \ {\rm b/s/Hz}.
\end{equation}

\emph{Example 1:}
In the QSSM system with $M=4$ and $L=4$, we have the data rate $R = 6$ b/s/Hz. Table I shows that $b_1b_2$ denotes the scatterer with the quadrature beam, $b_3b_4$ denotes the scatterer with the in-phase beam, and $b_5$ and $b_6$ represent the real part $x_\Re$ and imaginary part $x_\Im$ of the signal, respectively. We assume that the following input bits
\begin{equation}
\mathbf{b} = [\underbrace{0,1}_{{(b_1b_2)}},\underbrace{1,0}_{(b_3b_4)},\underbrace{1,1}_{(b_5b_6)}]
\end{equation}
are to be transmitted. Based on Table I, the first 2 bits modulate the active $2^\#$ scatterer index $l_\Re$ used to carry the real part of $\frac{1}{\sqrt{2}}$ corresponding to 5-th bit. Similarly, the 3-th and 4-th bits represent $3^\#$ scatterer activated by the in-phase beam, which carries the imaginary part of $\frac{1}{\sqrt{2}}$ corresponding to 6-th bit. To demonstrate the information in Table I clearly, we draw the QSSM constellation in Fig. \ref{constel}.

\subsection{QSSM Detection}

It is worth noting that when the receiver is equipped with a large number of antennas, the 40 MHz bandwidth can achieve centimeter-level positioning accuracy \cite{wymeersch}.
In the mmWave band, as the communication frequency increases, the positioning accuracy will become more precise.
The received signal of the QSSM scheme can be given as
\begin{equation}
\begin{aligned}
\mathbf{y}(k_1,k_2) = &\sqrt{P_s}\mathbf{H}
\left [ \begin{matrix}
\boldsymbol{a}_t(\theta_{k_1}^t)x_{\Re} +
\boldsymbol{a}_t(\theta_{k_2}^t)jx_{\Im}\\
\end{matrix} \right]
+\mathbf{n}
\\
=&\sqrt{P_s}\sum\limits_{k=1}^{L} \beta_l \boldsymbol{a}_r(\theta_l^r)\boldsymbol{a}_t^H(\theta_l^t)\boldsymbol{a}_t(\theta_{k_1}^t)x_{\Re} \\
&+ \sqrt{P_s}\sum\limits_{l=1}^{L} \beta_l \boldsymbol{a}_r(\theta_l^r)\boldsymbol{a}_t^H(\theta_l^t)\boldsymbol{a}_t(\theta_{k_2}^t)j x_{\Im} + \mathbf{n},
\end{aligned}
\end{equation}
where $P_s$ stands for the transmit power, $k_1, k_2 = 1, 2, \ldots, L$,
and $\mathbf n \in \mathbb{C}^{N_r\times 1}$ denotes the additive white Gaussian noise (AWGN) following $\mathcal{CN}(0,N_0\mathbf{I}_{N_r})$.
Note that vector $\mathbf n$ contains real and imaginary parts noises, each of part follows $\mathcal{CN}(0,{N_0}/{2})$.

To distinguish the direction of the received beam, we equip the receiver with a phase-shifted weight network to monitor the incoming wave direction of all scatterers in real time.
The phase-shifted weight network can be described as
$
\mathbf{R}^{r}=\left[\mathbf{R}^{r}_{1};\mathbf{R}^{r}_{2}\right]^{{H}},
$
where
$
\mathbf{R}^{r}_1=[\boldsymbol{a}_{{r}}(\theta_{1}^{r}), \cdots, \boldsymbol{a}_{{r}}(\theta_{k_1}^{r}), \cdots, \boldsymbol{a}_{{r}}(\theta_{L}^{r})] $ and
$
\mathbf{R}^{r}_2=[\boldsymbol{a}_{{r}}(\theta_{1}^{r}), \cdots, \boldsymbol{a}_{{r}}(\theta_{k_2}^{r}), \cdots, \boldsymbol{a}_{{r}}(\theta_{L}^{r})] $.
where every weight vector consists of phase shifters aiming at a specific receive beam.
Finally, the index of the scatterer can be obtained by judging the direction of the arrival beam. As such, the received signal after the weights of phase shifters can be expressed as
\begin{equation}
\begin{aligned}
&\mathbf{y}_{r}(k_1,k_2)\!
= \mathbf{R}_{r}\mathbf{y}(k_1,k_2)\\
=& \sqrt{P_s}\mathbf{R}_{r}\sum\limits_{k=1}^{L} \beta_l \boldsymbol{a}_r(\theta_l^r)\boldsymbol{a}_t^H(\theta_l^t)\boldsymbol{a}_t(\theta_{k_1}^t)x_{\Re}
\\
&+ \sqrt{P_s}\mathbf{R}_{r}\sum\limits_{l=1}^{L} \beta_l \boldsymbol{a}_r(\theta_l^r)\boldsymbol{a}_t^H(\theta_l^t)\boldsymbol{a}_t(\theta_{k_2}^t)j x_{\Im} + {n}_{r}\\
\!=\!& \sqrt{P_s}\boldsymbol{a}_{{r}}^{{H}}\left(\theta_{k_1}^r\right)  \boldsymbol{a}_r(\theta_{k_1}^r)\beta_{k_1}\boldsymbol{a}_t^H(\theta_{k_1}^t)\boldsymbol{a}_t(\theta_{k_1}^t)x_{\Re}
\\
&\!+\! \sqrt{P_s}\boldsymbol{a}_{{r}}^{{H}}\!\left(\theta_{k_2}^r\!\right)\!  \boldsymbol{a}_r(\theta_{k_2}^r)\beta_{k_2}\boldsymbol{a}_t^H(\theta_{k_2}^t)\boldsymbol{a}_t(\theta_{k_2}^t)j x_{\Im} \!+\! {n}_{r},
\end{aligned}
\end{equation}
where
\begin{equation}
{n}_{r}= \boldsymbol{a}_{{r}}^{{H}}\left(\theta_{k_1}^r\right)\mathbf{n}_{k_1} +
\boldsymbol{a}_{{r}}^{{H}}\left(\theta_{k_2}^r\right)\mathbf{n}_{k_2} \sim \mathcal{CN}(0, N_0),
\end{equation}
By utilizing Eq. (\ref{zhenj}), $\mathbf{y}_{r}$ can be calculated as
\begin{equation}
\begin{aligned}
\mathbf{y}_{r}(k_1,k_2)
=& \sqrt{P_s}\beta_{k_1}x_{\Re}
\!+\! \sqrt{P_s} \beta_{k_2}j x_{\Im} + {n}_{r}.
\end{aligned}
\end{equation}
Finally, the ML detection for the selected
scattering path can be characterized as
\begin{equation}\label{ml}
\begin{aligned}
&[\hat{k}_1, \hat{k}_2, \hat{x}_{\Re}, \hat{x}_{\Im}]\\
&= \arg\!\min\limits_{{k}_1, {k}_2, {x}_{\Re}, {x}_{\Im}}
|\mathbf{y}_{r}(k_1,k_2)\!-\!\sqrt{P_s}\left(\beta_{k_{1}}x_{\Re} \!+\! j\beta_{k_{2}}x_{\Im}\right)|^2,
\end{aligned}
\end{equation}
where $\hat{k}_1$ and $\hat{k}_2$ denote the indices of detected scatterers in the channel,
$\hat{x}_{\Re}$ and $\hat{x}_{\Im}$ stand for real and imaginary parts of the detected PSK/QAM signal.
\section{Performance Analysis}
In this section, we first give the closed-form expression on the pairwise error probability (PEP) of the QSSM scheme.
Then, we derive the asymptotic PEP expression of the QSSM scheme.
Finally, we provide the union upper bound expression of ABEP.

\subsection{Analytical PEP}
Based on Eq. (\ref{ml}),  the conditional PEP can be derived  as follows:
\begin{equation}\label{qf1}
\begin{aligned}
&P_r
=P\left(|\mathbf{y}_{r}(k_1,k_2)-\sqrt{P_s}\left(\beta_{k_{1}}x_{\Re} + j\beta_{k_{2}}x_{\Im}\right)|^2\right.\\ &\left.>|\mathbf{y}_{r}(k_1,k_2)-\sqrt{P_s}\left(\beta_{\hat{k}_{1}}\hat{x}_{\Re} + j\beta_{\hat{k}_{2}}\hat{x}_{\Im}\right)|^2\right)\\
&=P\!\left(\!-2\Re\left\{\!n_{r}\!\sqrt{P_s}\!\left[\!(\beta_{k_{1}}x_{\Re}\!-\!\beta_{\hat{k}_{1}}\!\hat{x}_{\Re}) \!+ \!j(\beta_{{k}_{2}}\!{x}_{\Im}\!-\!\beta_{\hat{k}_{2}}\!\hat{x}_{\Im})\right]\right\}
\right.\\ &\left.
-{P_s}\left[(\beta_{k_{1}}x_{\Re}-\beta_{\hat{k}_{1}}\hat{x}_{\Re}) + j(\beta_{{k}_{2}}{x}_{\Im}-\beta_{\hat{k}_{2}}\hat{x}_{\Im})\right]^2>0\right)\\
&=P(G>0),
\end{aligned}
\end{equation}
where
\begin{equation}
\begin{aligned}
G = \! &\!-2\Re\left\{\!n_{r}\!\sqrt{P_s}\!\left[(\beta_{k_{1}}x_{\Re}\!-\!\beta_{\hat{k}_{1}}\hat{x}_{\Re})
\!+ \!j(\beta_{{k}_{2}}{x}_{\Im}\!-\!\beta_{\hat{k}_{2}}\hat{x}_{\Im})\right]\right\}\\
&-{P_s}\left[(\beta_{k_{1}}x_{\Re}-\beta_{\hat{k}_{1}}\hat{x}_{\Re})
 + j(\beta_{{k}_{2}}{x}_{\Im}-\beta_{\hat{k}_{2}}\hat{x}_{\Im})\right]^2.
\end{aligned}
\end{equation}
Obviously, $G \sim \mathcal{N}(\mu_G  , \sigma_G^2)$, where $\mathcal{N}(\cdot,\cdot)$ denotes the real Gaussian distribution and the mean and variance are expressed as
\begin{equation}
\begin{aligned}
&\mu_G= -{P_s}\left|(\beta_{k_{1}}x_{\Re}-\beta_{\hat{k}_{1}}\hat{x}_{\Re}) + j(\beta_{{k}_{2}}{x}_{\Im}-\beta_{\hat{k}_{2}}\hat{x}_{\Im})\right|^2, \\
&\sigma_G^2\!= \! -2N_0{P_s}\left|(\beta_{k_{1}}x_{\Re}\!-\!\beta_{\hat{k}_{1}}\hat{x}_{\Re}) \!+ \!j(\beta_{{k}_{2}}{x}_{\Im}\!-\!\beta_{\hat{k}_{2}}\hat{x}_{\Im})\right|^2.
\end{aligned}
\end{equation}
Since $P_r = P(G>0)=P(-\mu_G/\sigma_G)$, Eq. (\ref{qf1}) can be rewritten as
\begin{equation}\label{qfunc}
\begin{aligned}
P_r
&=Q\left(\sqrt{\frac{{P_s}\left|(\beta_{k_{1}}x_{\Re}-\beta_{\hat{k}_{1}}\hat{x}_{\Re}) + j(\beta_{{k}_{2}}{x}_{\Im}-\beta_{\hat{k}_{2}}\hat{x}_{\Im})\right|^2}{2N_0}}\right)\\
&=Q\left(\sqrt{\frac{{\rho}\eta}{2}}\right),
\end{aligned}
\end{equation}
where
$Q(\cdot)$ denotes the $Q$ function,
$\rho ={P_s}/{N_0}$ {indicates signal-to-noise ratio (SNR)}, and $\sqrt{\eta} ={{(\beta_{k_{1}}x_{\Re}-\beta_{\hat{k}_{1}}\hat{x}_{\Re}) + j(\beta_{{k}_{2}}{x}_{\Im}-\beta_{\hat{k}_{2}}\hat{x}_{\Im})}}$.
Since zero mean of $\sqrt{\eta}$ and orthogonal beams of $k_1$ and $k_2$,
the variance of $\sqrt{\eta}$ can be obtained as
\begin{figure}[t]
  \centering
  \includegraphics[width=6.3cm]{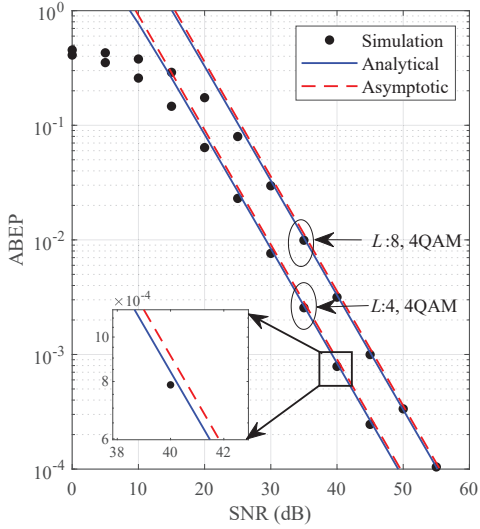}\\
  \caption{\small{ABEP analytical derivation verification.}}\label{f1}
\end{figure}
\begin{equation}\label{eta}
\bar{\eta}\!=\!\\
\begin{cases}
|x_{\Re}-\hat{x}_{\Re}|^2+|x_{\Im}-\hat{x}_{\Im}|^2,\!
&{ {\beta}_{k_{1}}\! = \!{ \beta}_{\hat{k}_{1}}, {\beta}_{k_{2}}\! =\! { \beta}_{\hat{k}_{2}}}\\
|x_{\Re}|^2 + |\hat{x}_{\Re}|^2+|x_{\Im}-\hat{x}_{\Im}|^2,\!
&{ {\beta}_{k_{1}} \!\neq \!{ \beta}_{\hat{k}_{1}}, {\beta}_{k_{2}}\! =\!{ \beta}_{\hat{k}_{2}}},\\
|x_{\Re}-\hat{x}_{\Re}|^2+|x_{\Im}|^2 + |\hat{x}_{\Im}|^2,\!
& { {\beta}_{k_{1}} \!=\! { \beta}_{\hat{k}_{1}}, {\beta}_{k_{2}}\! \neq \!{ \beta}_{\hat{k}_{2}}},\\
|x_{\Re}|^2 + |\hat{x}_{\Re}|^2+|x_{\Im}|^2 + |\hat{x}_{\Im}|^2,\!
& { {\beta}_{k_{1}} \!\neq \!{\beta}_{\hat{k}_{1}}, {\beta}_{k_{2}} \!\neq \!{ \beta}_{\hat{k}_{2}}}.
\end{cases}
\end{equation}
Since the signal power is reflected at the receive SNR, we need to normalize $\eta$.
Let us define $\gamma = \frac{\eta}{{\bar{\eta}}}$, the probability density function of the central Chi-square distribution with two degrees of freedom can be evaluated as \cite{Kay1998}
\begin{equation}\label{pdf}
\begin{aligned}
f(\gamma) &= \frac{1}{2}\exp\left(-\frac{\gamma}{2}\right).
\end{aligned}
\end{equation}
Combining (\ref{qfunc}) and (\ref{pdf}), the unconditional PEP can be characterized as
\begin{equation}\label{pr1}
\begin{aligned}
\bar{P}_r &=  \frac{1}{2}\int_0^\infty \exp\left(-\frac{\gamma}{2}\right)Q\left(\sqrt{\frac{\rho \bar{\eta}\gamma}{2}}\right)d\gamma.
\end{aligned}
\end{equation}
Substituting  $Q(\gamma) = \frac{1}{\pi}\int_0^{\frac{\pi}{2}}\exp\left(-\frac{\gamma^2}{2\sin^2\theta}\right)d\theta$ into (\ref{pr1}), the unconditional PEP can be calculated as
\begin{equation}
\begin{aligned}
\bar{P}_r
&=\frac{1}{2\pi}\int_0^\infty \int_0^{\frac{\pi}{2}}\exp\left(-\frac{{2\sin^2\theta}+\rho \bar{\eta}}{4\sin^2\theta}\gamma\right)d\theta d\gamma.
\end{aligned}
\end{equation}
Exchanging the order of integration of $\theta$ and $\gamma$, we have
\begin{equation}\label{rtheta}
\begin{aligned}
\bar{P}_r
&=\frac{1}{2\pi}\int_0^{\frac{\pi}{2}}\int_0^\infty \exp\left(-\frac{{2\sin^2\theta}+\rho \bar{\eta}}{4\sin^2\theta}\gamma\right)d\gamma d\theta.
\end{aligned}
\end{equation}
Addressing the inner integral, Eq. (\ref{rtheta}) can be recast as
\begin{equation}
\begin{aligned}
\bar{P}_r
&=\frac{1}{\pi}\int_0^{\frac{\pi}{2}} \left(\frac{\sin^2\theta}{{\sin^2\theta}+\frac{\rho \bar{\eta}}{2}}\right)d\theta.
\end{aligned}
\end{equation}
After some algebraic operations, we can obtain
\begin{equation}\label{pepfin}
\begin{aligned}
\bar{P}_r
&=\frac{1}{2}\left(1-\sqrt{\frac{1}{1+\frac{2}{\rho \bar{\eta}}}}\right).
\end{aligned}
\end{equation}
\begin{figure}[t]
  \centering
  \includegraphics[width=6.3cm]{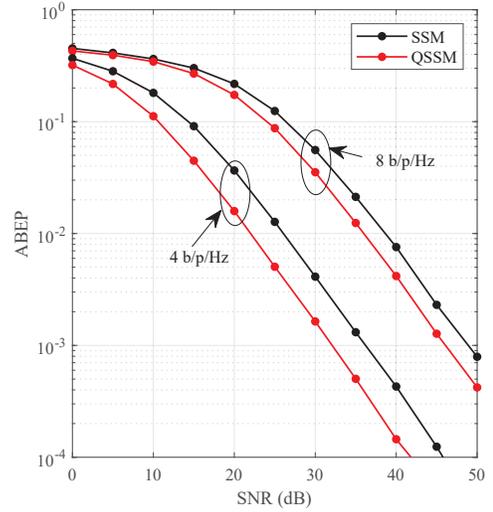}\\
  \caption{\small{Comparisons with SSM.}}\label{f2}
\end{figure}
\subsection{Asymptotic PEP}
To clarify the trend of the PEP and to gain more valuable insights, we derive the asymptotic PEP of the QSSM in this subsection.

Substituting  $Q(\gamma) \approx \frac{1}{12}\exp\left(-\frac{\gamma^2}{2}\right)+\frac{1}{4}\exp\left(-\frac{2\gamma^2}{3}\right)$ into (\ref{pr1}), we can obtain the loose upper bound of $\bar{P}_r$ as follows:
\begin{equation}\label{pr22}
\begin{aligned}
\bar{P}_r
\!=\!&\frac{1}{24}\!\int_0^\infty \!\exp\left(\!-\frac{2\!+\!\rho \bar{\eta}}{4}\gamma\right)\! d\gamma
\!+\! \frac{1}{8}\int_0^\infty\! \exp\left(\!-\frac{3\!+\!2\rho \bar{\eta}}{6}\gamma\!\right) \! d\gamma.
\end{aligned}
\end{equation}
In high SNR regions, the constant term added to $\rho\bar{\eta}$ can be ignored.
Consequently, Eq. (\ref{pr22}) can be evaluated as
\begin{equation}\label{asy11}
\begin{aligned}
\lim\limits_{\rho \to \infty}\bar{P}_r
=&\frac{1}{24}\int_0^\infty \exp\left(-\frac{\rho \bar{\eta}}{4}\gamma\right) d\gamma + \frac{1}{8}\int_0^\infty \exp\left(-\frac{\rho \bar{\eta}}{3}\gamma\right) d\gamma \\
=&
\frac{13}{24\rho \bar{\eta}}.
\end{aligned}
\end{equation}

\begin{figure*}[t]
\begin{equation}\label{abep}
\begin{aligned}
{\rm ABEP} \leq & \sum_{k_1 = 1}^L \sum_{{k_2}=1}^L \sum_{m=1}^{M}  \sum_{\hat{k} = 1}^L \sum_{{\hat{k}_2}=1}^L \sum_{\hat{m}=1}^{M}\frac{N([k_1,k_2,m] \to [\hat{k}_1,\hat{k}_2,\hat{m}])\bar{P}_r
}{{{L^2M  \log _{2}(L^2M )}}}.
\end{aligned}
\end{equation}
\hrulefill
\end{figure*}

\subsection{ABEP}
With the unconditional PEP and asymptotic PEP in (\ref{pepfin}) and (\ref{asy11}), we provide the corresponding ABEP in (\ref{abep}),
shown at the top of next page,
where
$N([k_1,k_2,m] \to [\hat{k}_1,\hat{k}_2,\hat{m}])$ denotes the Hamming distance between symbol $[k_1,k_2,m]$ {and} $[\hat{k}_1,\hat{k}_2,\hat{m}]$.
Besides, $\log_2(L^2M)$ represents the total number of constellation points in the QSSM constellation diagram.

\section{Numerical Results}
In this section, we evaluate the ABEP of the QSSM scheme via the simulations and derivation results.
Note that the simulation for each value is obtained by averaging $10^6$ random numbers.
According to \cite{ding2017ssm}, we  set $N_t =32$,$N_t =32$, and $d_t = d_r = {\lambda}/{2}$.

In Fig. \ref{f1}, the analytical and asymptotic results are presented and compared with the simulation values, where we employ $L=4$ and $L=8$ QSSM systems with 4QAM to achieve $ {\rm 6 \ b/s/Hz}$ and $ {\rm 8 \ b/s/Hz}$ data rates, respectively.
In the extensive and practical SNR region, the analytical and simulation results match well,
which validates the derivation results of ABEP. However, in the low SNR region, there is a noticeable mismatch.
This is because the derived results represent the upper bounds rather than exact ABEP.
It can be observed that the difference between analytical and asymptotic results is small and almost negligible in the high SNR region.
To some extent, we can replace Eq. (\ref{asy11}) with Eq. (\ref{pepfin}) to characterize the ABEP of the QSSM scheme.

\begin{figure}[t]
  \centering
  \includegraphics[width=6.3cm]{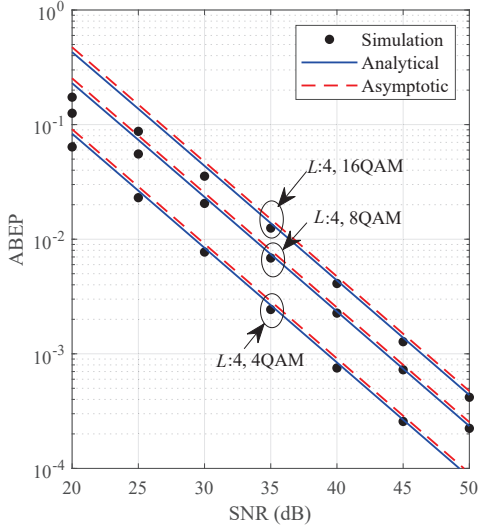}\\
  \caption{\small{ABEP under different data rates.}}\label{f3}
\end{figure}

In Fig. \ref{f2}, compared with the conventional SSM scheme, we plot the ABEP of the QSSM scheme under the same spectral efficiency.
As the data rate is 4 b/p/Hz, corresponding parameters are set as $L = 2$ and 4QAM, while the data rate is 8 b/p/Hz, we let $L = 4$ and 16QAM.
It is evident from Fig. \ref{f2} that QSSM provides better ABEP performance than SSM. Specifically, for 4 b/p/Hz, the QSSM exhibits an SNR gain of about 3.6 dB when ABEP is equal to $10^{-4}$. In addition, for 4 b/p/Hz, the QSSM achieves an SNR gain of about 3 dB at ABEP = $10^{-3}$.

Fig. \ref{f3} shows the simulations, analytical, and asymptotic results for the ABEP versus SNR for $L = 4$ and different modulation orders.
It can be seen from Fig. \ref{f3} that both the analytical upper bound and the asymptotic results are tight within the SNR range of Fig. \ref{f3}.
Furthermore, the 4QAM corresponding to QSSM system achieves better ABEP performance than that of 8QAM and 16QAM. For example,
when ABEP = $10^{-3}$, compared to 4QAM, the systems of 8QAM and 16QAM require 4.3 dB and 7.1 dB SNRs, respectively.

\section{Conclusions}

In this letter, we considered a novel QSSM scheme aided MIMO architecture for mmWave transmission systems.
Based on ML detection, the closed-form expressions of ABEP on the QSSM scheme was derived over the Saleh-Valenzuela channel.
To get more valuable sights, the asymptotic ABEP expression of QSSM scheme are also provided in this letter.
Finally, Monte Carlo simulations validated our analysis and showed that the ABEP performance of QSSM system is better than that of conventional SSM system with different system configurations.
We hope that the insights provided in this work will be helpful for future research in QSSM.
In future work, QSSM schemes under non-ideal conditions, such as inter-beam interference and imperfect a priori channel knowledge, remain an attractive and open research topic.

\end{document}